\documentclass{Interspeech}

\interspeechcameraready

\title{Factorized RVQ-GAN For Disentangled Speech Tokenization}

\author[affiliation={1^{\star}}]{Sameer}{Khurana}
\author[affiliation={2^{\star}}]{Dominik}{Klement}
\author[affiliation={3^{\star}}]{Antoine}{Laurent}
\author[affiliation={^{\spadesuit}}]{Dominik}{Bobos}
\author[affiliation={^{\spadesuit}}]{Juraj}{Novosad}
\author[affiliation={^{\spadesuit}}]{Peter}{Gazdik}
\author[affiliation={^{\diamondsuit}}]{Ellen}{Zhang}
\author[affiliation={^{\heartsuit}}]{Zili}{Huang}
\author[affiliation={^{\heartsuit}}]{Amir}{Hussein}
\author[affiliation={^{\clubsuit}}]{Ricard}{Marxer}
\author[affiliation={1}]{Yoshiki}{Masuyama}
\author[affiliation={1}]{Ryo}{Aihara}
\author[affiliation={1}]{Chiori}{Hori}
\author[affiliation={1}]{François}{G. Germain}
\author[affiliation={1}]{Gordon}{Wichern}
\author[affiliation={1}]{Jonathan}{Le Roux}

\affiliation{}{Mitsubishi Electric Research Laboratories (MERL)}{Cambridge, MA, USA}
\affiliation{Speech@FIT}{Brno University of Technology}{Czech Republic \quad $^3$LIUM University, France}
\email{sameerkhurana10@gmail.com, xkleme15@vutbr.cz, leroux@merl.com}

\usepackage{comment}
\usepackage{graphicx}
\usepackage{caption}
\usepackage{cite}
\usepackage{subcaption}

\makeatletter
\def\bstctlcite{\@ifnextchar[{\@bstctlcite}{\@bstctlcite[@auxout]}}
\def\@bstctlcite[#1]#2{\@bsphack
  \@for\@citeb:=#2\do{%
    \edef\@citeb{\expandafter\@firstofone\@citeb}%
    \if@filesw\immediate\write\csname #1\endcsname{\string\citation{\@citeb}}\fi}%
  \@esphack}
\makeatother 

\keywords{RVQ, GAN, Audio Codec, Speech Tokenization}

\begin{document}
\bstctlcite{IEEEexample:BSTcontrol} 

\maketitle

\begin{abstract}
We propose Hierarchical Audio Codec (HAC), a unified neural speech codec that factorizes its bottleneck into three linguistic levels—acoustic, phonetic, and lexical—within a single model. HAC leverages two knowledge distillation objectives: one from a pre-trained speech encoder (HuBERT) for phoneme-level structure, and another from a text-based encoder (LaBSE) for lexical cues. Experiments on English and multilingual data show that HAC’s factorized bottleneck yields disentangled token sets: one aligns with phonemes, while another captures word-level semantics. Quantitative evaluations confirm that HAC tokens preserve naturalness and provide interpretable linguistic information, outperforming single-level baselines in both disentanglement and reconstruction quality. These findings underscore HAC’s potential as a unified discrete speech representation, bridging acoustic detail and lexical meaning for downstream speech generation and understanding tasks.
\end{abstract}

\begingroup
\makeatletter
\renewcommand\@makefnmark{}
\renewcommand{\@makefntext}[1]{\noindent#1}
\makeatother
\footnotetext[0]{$^{\star}$Major Contributors. $^{\spadesuit}$Phonexia, Czech Republic. $^{\diamondsuit}$Massachusetts Institute of Technology, USA. $^{\heartsuit}$CLSP, Johns Hopkins University, USA. $^{\clubsuit}$Univ.\ de Toulon, Aix Marseille Univ., CNRS, LIS, Toulon, France.}
\endgroup

\section{Introduction}
Neural speech codecs (NSCs) are a family of neural network architectures that convert speech signals into discrete token representations~\cite{zeghidour2021soundstream,defossez2023high,kumar2023highfidelity,ESPnetCodec}. These discrete tokens can then be leveraged in various downstream tasks, ranging from spoken language modeling~\cite{borsos2023audiolm} and speech-to-speech translation~\cite{lee-etal-2022-direct} to text-to-speech synthesis~\cite{valle} and speech understanding in large language models. Broadly, NSCs can be classified into phonetic (P-NSC) and acoustic (A-NSC) approaches.

A P-NSC follows a two-stage pipeline: (1) a pre-trained transformer encoder (e.g., HuBERT \cite{hsu2021hubert}), trained via self-supervised learning (SSL), outputs contextual acoustic frame embeddings; 
 and (2) those embeddings, selected from the layer that performs best on a phoneme recognition task, are quantized using $k$-means vector quantization (VQ) to produce discrete tokens~\cite{ASR2}.
 Because these tokens align closely with underlying phoneme labels, P-NSCs excel at tasks requiring high-level linguistic structure \cite{lee-etal-2022-direct, lakhotia2021generative}. However, speech generated from P-NSCs tends to sound robotic and lacks speaker diversity \cite{borsos2023audiolm}.

By contrast, A-NSCs aim for high-fidelity speech reconstruction through a low-bitrate compression model, often based on the residual VQ-generative adversarial network (RVQ-GAN) framework. An encoder maps input speech to acoustic frame embeddings 
which are then quantized into discrete token sequences 
across multiple VQ layers. A decoder reconstructs the speech from these tokens, preserving detailed acoustic nuances, resulting in more natural-sounding speech with broad speaker variation. Nonetheless, because A-NSCs focus on fine-grained acoustic details, they often lack coherent linguistic and grammatical structure. Prominent examples of A-NSCs include SoundStream \cite{zeghidour2021soundstream}, Encodec \cite{defossez2023high}, and Descript Audio Codec \cite{kumar2023highfidelity}.

Recent work, such as SpeechTokenizer \cite{zhang2024speechtokenizer}, addresses the limitations of standalone P-NSCs and A-NSCs by combining phonetic and acoustic tokens within a single framework. However, these two-level solutions omit an important lexical representation, which captures word-level or subword-level semantic and syntactic information. In contrast, our proposed Hierarchical Audio Codec (HAC) introduces this third level of abstraction, lexical, alongside phonetic and acoustic tokens, enabling it to jointly model higher-level linguistic structure, mid-level phonetic details, and fine-grained acoustic nuances. Learning this extra lexical layer is particularly valuable for downstream tasks such as spoken language modeling, speech-to-speech translation, and voice-enabled question answering, where a richer, word-oriented representation can substantially improve semantic coherence and contextual accuracy. By disentangling the token space across these three levels, HAC combines the strengths of existing two-level models with a deeper linguistic understanding, all within a unified architecture that eliminates the need for external token merging.

\section{Hierarchical Audio Codec (HAC)}
\label{sec:hac}
\begin{figure}
    \centering
    \includegraphics[width=0.99\linewidth]{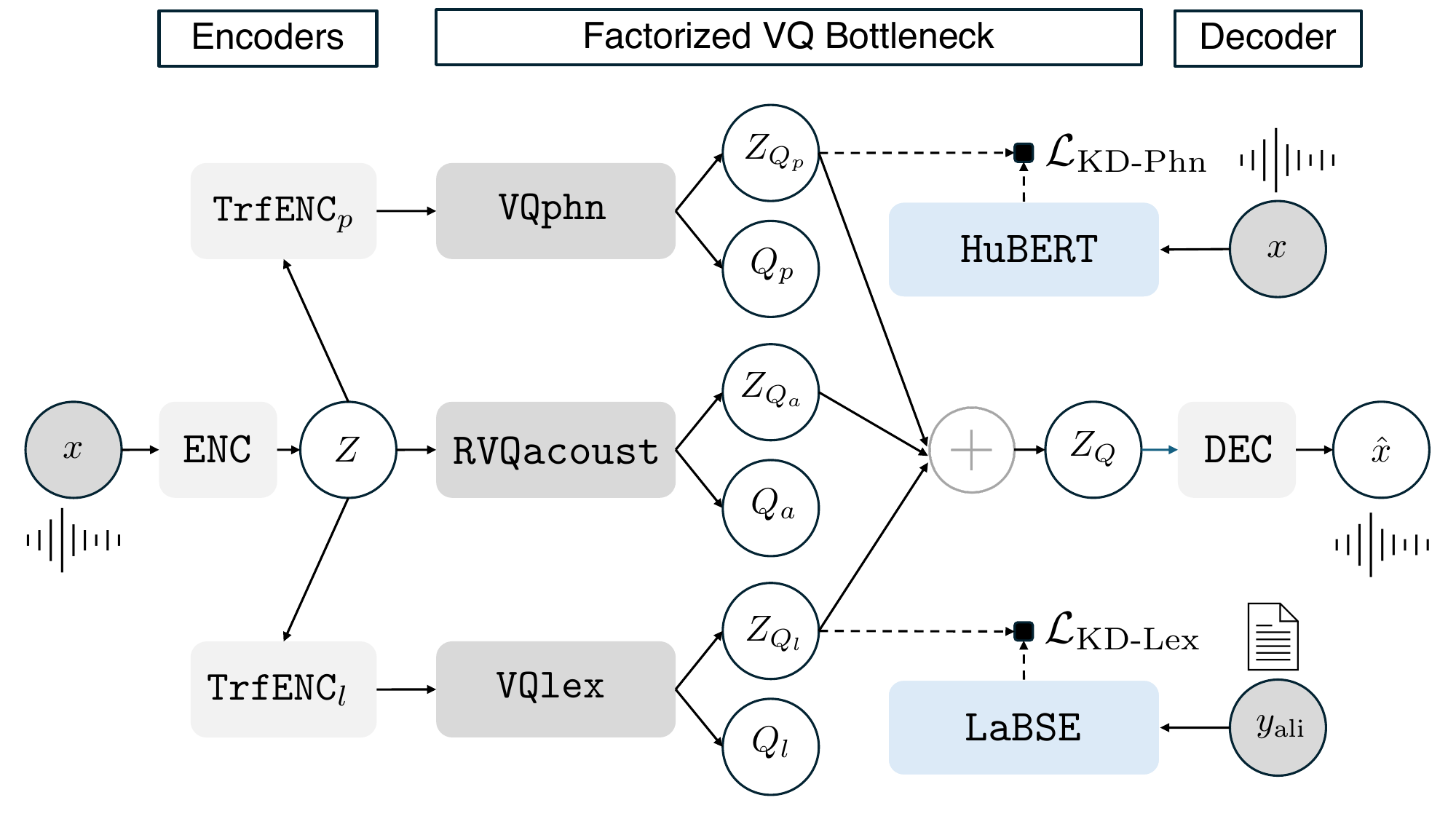}
    \caption{Diagram of our proposed Hierarchical Audio Codec (HAC). HAC encodes the input speech signal $x$ into a multi-level set of disentangled discrete tokens (lexical $Q_l$, phonetic $Q_p$, and acoustic $Q_a$) capturing distinct aspects of the audio.}
    \vspace{-.3cm}
    \label{fig:hac}
\end{figure}
HAC, illustrated in Fig.~\ref{fig:hac}, consists of a down-sampling CNN encoder (${\tt ENC}$), two transformer encoders (${\tt TrfENC}_{p,l}$), a factorized bottleneck consisting of three VQ modules (${\tt VQphn}$, ${\tt RVQacoust}$, and ${\tt VQlex}$), and an up-sampling CNN decoder (${\tt DEC}$). 
HAC is trained using tuples $(x, y_{\text{ali}})$, where $x \in \mathbb{R}^{T}$ is a speech utterance and $y_{\text{ali}}$ is its force-aligned text transcript. HAC maps $x$ to $\hat{x}$, a reconstruction of $x$ through the following steps:
\begin{align*}
    Z &= {\tt ENC}(x), \\
    Z_{Q_a}, Q_a &= {\tt RVQacoust}(Z), \\
    Z_{Q_p}, Q_p &= {\tt VQphn}({\tt TrfENC}_p(Z)), \\
    Z_{Q_l}, Q_l &= {\tt VQlex}({\tt TrfENC}_l(Z)), \\
    Z_Q &= Z_{Q_p} + Z_{Q_a} + Z_{Q_l}, \\
    \hat{x} &= {\tt DEC}(Z_Q),
\end{align*}
where $Z,\ Z_{Q_a},\ Z_{Q_p},\ Z_{Q_l} \in \mathbb{R}^{F\times D}$, and $Z_{Q_a},\ Z_{Q_p}$, and $Z_{Q_l}$ are the codebook entries corresponding to token sets $Q_a \in \{1, \ldots, A\}^{F\times N}$, $Q_p \in \{1, \ldots, P\}^{F}$, and $Q_l \in \{1, \ldots, L\}^{F}$, respectively. Here, $F$ is the number of acoustic frames, and $D$ is the frame embedding dimension. To maximize codebook utilization, the VQ layers follow the low-dimensional code lookup process described in \cite{kumar2023highfidelity}. ${\tt RVQacoust}$ module performs residual vector quantization (RVQ) and consists of $N$ VQ layers, each with a codebook size of $A$. ${\tt VQphn}$ and ${\tt VQlex}$ consist of a single VQ layer with codebook sizes $P$ and $L$, respectively.

HAC is trained in an adversarial framework, where the HAC generator is paired with a discriminator as described in~\cite{kumar2023highfidelity}. The overall training objective includes: 1) a frequency-domain reconstruction loss to ensure faithful spectral recovery, 2) an adversarial loss to encourage natural-sounding outputs, and 3) codebook learning losses to update the codebook entries. These objectives are described in detail in \cite{kumar2023highfidelity}. To ensure that each token set encodes the intended type of information, we introduce knowledge distillation (KD) losses on the phonetic and lexical bottlenecks: 1) $\mathcal{L}_{\text{KD-Phn}}$ encourages $Q_p$ to represent phoneme-level features, and 2) $\mathcal{L}_{\text{KD-Lex}}$ encourages $Q_l$ to represent word-level (lexical) information. By providing the decoder with phonetic ($Z_{Q_p}$) and high-level lexical ($Z_{Q_l}$) information, the acoustic bottleneck ${\tt RVQacoust}$ is free to focus on fine-grained acoustic details, critical for high-fidelity signal reconstruction. 

Following \cite{zhang2024speechtokenizer}, we compute the KD losses as follows:
\begin{align*}
    \tilde{Z}_{Q_p} &= Z_{Q_p}A_p, \quad \tilde{Z}_{Q_l} = Z_{Q_l}A_l,\\
    \mathcal{L}_{\text{KD-Phn}}&= -\frac{1}{D^{\prime}}\sum\limits_{d=1}^{D^{\prime}}\log (\sigma(\operatorname{cos\_sim}(\tilde{Z}_{Q_p}[:, d], Z_{\text{hubert}}[:, d]))),\\
    \mathcal{L}_{\text{KD-Lex}}&= -\frac{1}{D^{\prime\prime}}\sum\limits_{d=1}^{D^{\prime\prime}}\log (\sigma(\operatorname{cos\_sim}(\tilde{Z}_{Q_l}[:, d], Z_{\text{labse}}[:, d]))),\\
    Z_{\text{hubert}} &= {\tt Avg}({\tt HuBERT}(x)),\ \ \ \ Z_{\text{labse}} = {\tt Avg}({\tt LaBSE}(y_{\text{ali}})),
\end{align*}
where $D^{\prime}$ and $D^{\prime\prime}$ are the embedding dimensionalities of HuBERT and LaBSE, respectively, $\operatorname{cos\_sim}(\cdot)$ is the cosine similarity, $\sigma(\cdot)$ is the sigmoid function, $A_p$ and $A_l$ are projection matrices for dimension matching, and ${\tt Avg(\cdot)}$ averages representations over all layers of HuBERT or LaBSE \cite{feng-etal-2022-language}.

While we explore other settings in Section~\ref{sec:exp}, our best-performing HAC model has the following hyperparameters: 1) Training input: each speech recording $x$ is 3.8 seconds long and has a 16 kHz sampling rate; 2) Downsampling factor: ${\tt ENC}$ reduces the input time resolution by a factor of 320; 3) Frame embedding dimensionality: $D=1024$; 4) Phonetic and lexical codebooks: both ${\tt VQphn}$ and ${\tt VQlex}$ use a single VQ layer with codebook size 16,384 (14-bit), and codebook entries are 128-dimensional; and 5) Acoustic codebook: ${\tt RVQacoust}$ has $N=7$ VQ layers, each with codebook size $A=1024$, and each codebook entry is 8-dimensional. Each acoustic frame is represented by $9$ tokens (the $7$ acoustic tokens plus $1$ phonetic token and $1$ lexical token). The ${\tt ENC}$ and ${\tt DEC}$ follow the same CNN architecture described in \cite{kumar2023highfidelity}. ${\tt TrfENC}_{p,l}$ has $4$ layers, $8$ attention heads, model embedding dimensionality of $768$, and feed-forward dimensionality $3072$. Layer-normalization is applied to each layer's input, and the encoder has learnable convolutional positional embeddings \cite{baevski2020wav2vec}.

HAC is optimized on eight A40 GPUs with a total batch size of 60 seconds. We use the AdamW optimizer with a learning rate 1e-4, $\beta_1 = 0.8$, and $\beta_2 = 0.9$. We train the model for 400K iterations and decay the learning rate at every step with $\gamma = 0.999996$.

\section{Experiments}
\label{sec:exp}
\begin{figure*}
    \centering
    \includegraphics[width=0.75\linewidth]{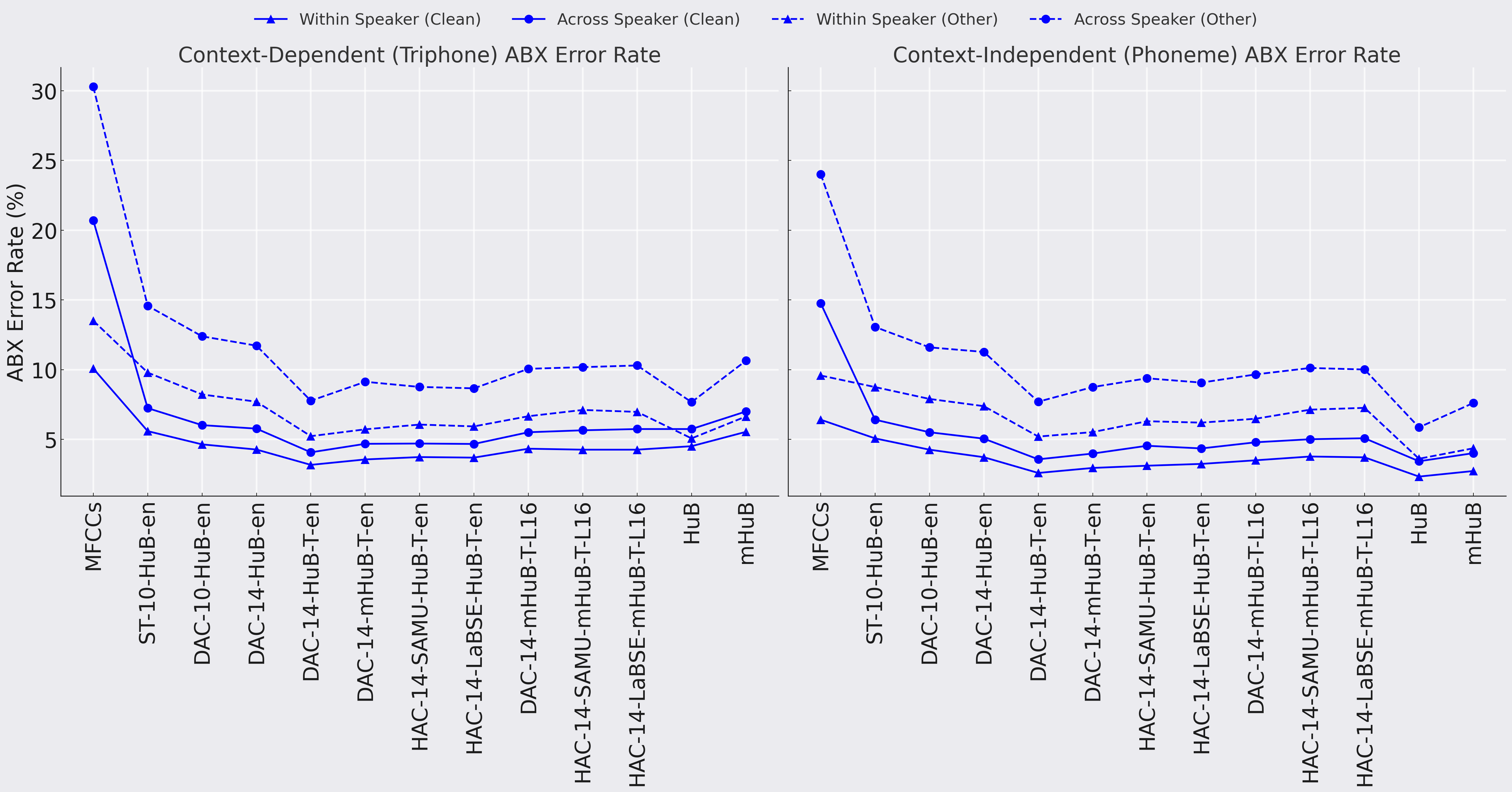}
    \vspace{-.1cm}
    \caption{ABX error rate across different models and scenarios.}
    \vspace{-.45cm}
    \label{fig:abx}
\end{figure*}
We train English-language models on the LibriSpeech dataset \cite{7178964}, consisting of 960 hours of transcribed English speech, and multilingual models on the VoxPopuli dataset \cite{wang-etal-2021-voxpopuli}, which includes 1.7K hours of transcribed speech across 16 languages. For evaluation, we use forced-aligned test sets from LibriSpeech and Multilingual LibriSpeech (MLS) \cite{pratap20_interspeech}. The MLS corpus contains eight languages: English (EN), French (FR), German (DE), Italian (IT), Polish (PL), Portuguese (PT), Spanish (ES), and Dutch (NL). We obtain forced alignments using the Montreal Forced Aligner \cite{mcauliffe17_interspeech}. We train this aligner for languages other than English on a subset of each respective MLS training set. 

Below, we summarize the models trained in this work. Each model name encodes its key features, including codebook size, the teacher model for KD, and whether a transformer encoder is employed.

\textbf{ST-10-HuB-en (Baseline):} This is our English (en) baseline, 
SpeechTokenizer (ST) \cite{zhang2024speechtokenizer}. It is based on an RVQ-GAN framework trained with a generative objective and the phonetic KD loss ($\mathcal{L}_{\text{KD-Phn}}$) described in Section~\ref{sec:hac}. ST's RVQ bottleneck consists of $9$ VQ layers. The model distills from HuBERT-Base by matching the first VQ layer’s quantized frame embeddings to the averaged HuBERT-Base embeddings. Each VQ layer has a 10-bit codebook of dimensionality $1024$. ST extends~Encodec \cite{defossez2023high} (an earlier RVQ-GAN for audio compression) by adding the Phoneme level KD loss to Encodec’s original generative and codebook learning objectives. Like Encodec, ST updates its codebooks via exponential moving average (EMA) and periodically re-initializes them to maximize utilization. To have a fair comparison with other models, we train the baseline using our codebase, instead of using the publicly available checkpoint.

\textbf{DAC-10-HuB-en:} This model adds the phonetic KD loss to DAC \cite{kumar2023highfidelity}, an improved version of Encodec that uses low-dimensional code lookups for improved codebook utilization and reconstruction. Like the baseline, it has $9$ 10-bit VQ layers but has a codebook dimensionality of $8$, much lower than the baseline. Since a very low-dimensional codebook is less ideal for KD, we increase the first VQ layer’s dimensionality to $128$ while keeping the remaining layers at $8$ dimensions. Furthermore, unlike ST-10-HuB-en, where the phonetic VQ layer (the KD student) is part of the RVQ module, in DAC-10-HuB-en, we factor out this layer, leading to slightly improved token quality and reconstruction performance. \textbf{DAC-14-HuB-en:} This variant extends DAC-10-HuB-en by increasing the phonetic VQ layer’s codebook size from 10-bit to 14-bit, enabling a larger vocabulary of phoneme-level tokens. \textbf{DAC-14-HuB-T-en:} Building on DAC-14-HuB-en, this model adds a transformer encoder (${\tt TrfENC}_p$ described in Section~\ref{sec:hac}) before the factored-out phonetic VQ layer, allowing the VQ layer to capture richer context before quantization. \textbf{DAC-14-mHuB-T-en:} This version uses multilingual HuBERT (mHuBERT) \cite{boito2024mhubert} as the teacher for phonetic KD loss. It follows the same architecture as DAC-14-HuB-T-en but replaces HuBERT-Base with a multilingual variant. \textbf{DAC-14-mHuB-T-L16:} A multilingual DAC model that is trained on both VoxPopuli and LibriSpeech. It retains the 14-bit phonetic VQ layer and uses mHuBERT, matching DAC-14-mHuB-T-en’s setup, but extends training across 16 languages.

\begin{figure}
    \centering
    \includegraphics[width=0.99\linewidth]{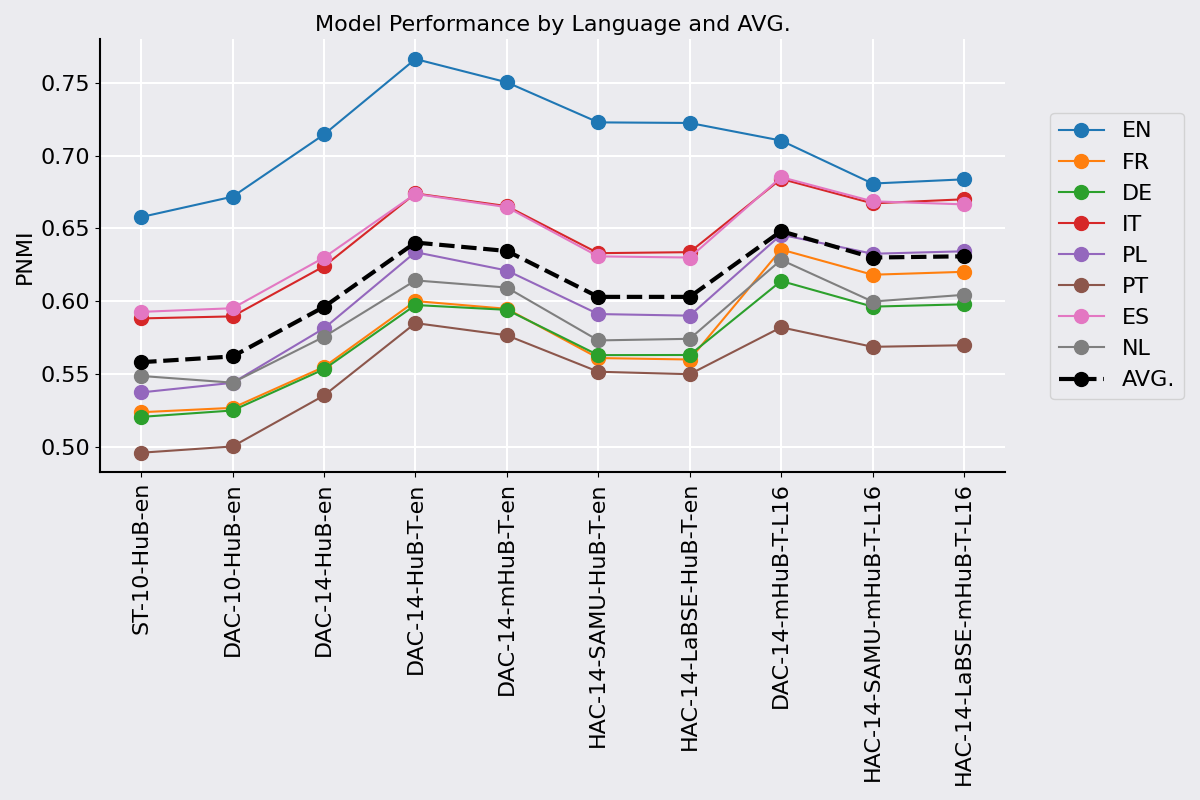}
    \caption{Phoneme Normalized Mutual Information (PNMI) across different models and languages.}
    \vspace{-.1cm}
    \label{fig:pnmi_multi}
    \vspace{-.3cm}
\end{figure}
\textbf{HAC-14-SAMU-HuB-T-en:} This is the Hierarchical Audio Codec (HAC) variant with $9$ VQ layers, of which two are factored out of the RVQ module. Both factored-out layers have 14-bit codebooks of dimensionality $128$; the remaining layers have 10-bit codebooks of dimension $8$. For the lexical-level KD loss ($\mathcal{L}_{\text{KD-Lex}}$), we use SAMU-XLS-R (SAMU) \cite{khurana2022samu} as the teacher. SAMU is a language-agnostic semantic speech encoder obtained by distilling LaBSE into a speech model. SAMU removes the need for forced-aligned transcripts when computing the lexical KD loss. Meanwhile, phonetic KD loss still uses HuBERT-Base. \textbf{HAC-14-LaBSE-HuB-T-en:} Identical to HAC-14-SAMU-HuB-T-en, except the lexical KD loss is computed directly from LaBSE embeddings that correspond to force-aligned transcripts rather than from SAMU. \textbf{HAC-14-SAMU-mHuB-T-L16:} A multilingual HAC model, extending HAC-14-SAMU-HuB-T-en to 16 languages. It uses mHuBERT for phoneme-level KD and SAMU for lexical-level KD. \textbf{HAC-14-LaBSE-mHuB-T-L16:} The multilingual version of HAC-14-LaBSE-HuB-T-en. It uses LaBSE for lexical-level KD and mHuBERT for phoneme-level KD.

\begin{figure}
    \centering
    \includegraphics[width=0.99\linewidth]{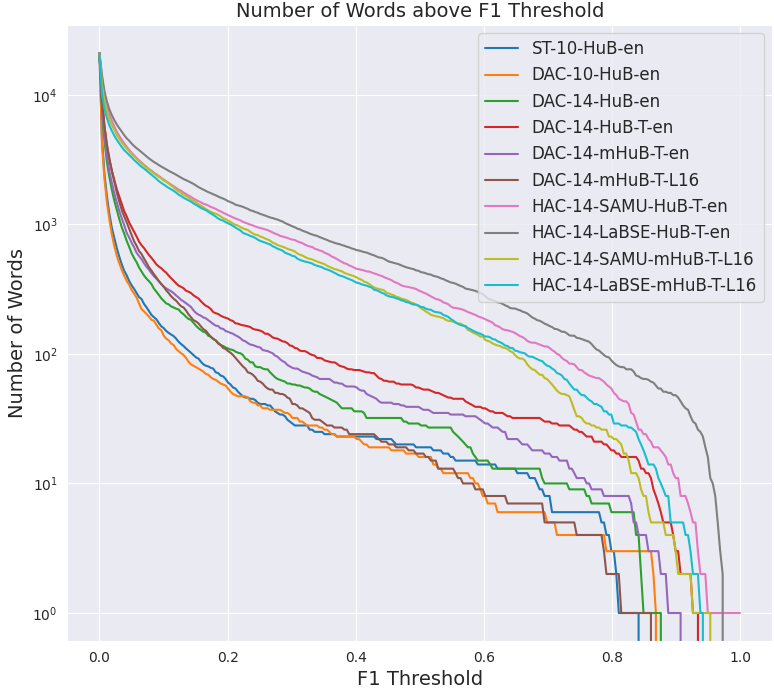}
    \vspace{-.15cm}
    \caption{F1 Scores of VQ tokens when treated as word detectors on the LibriSpeech corpus.}
    \label{fig:word_detects}
    \vspace{-.45cm}
\end{figure}

Figures \ref{fig:abx} and \ref{fig:pnmi_multi} evaluate how well the phonetic VQ layer (attached to $\mathcal{L}_{\text{KD-Phn}}$) captures phoneme-level information.

Figure~\ref{fig:abx} compares various models on the ABX discrimination task \cite{phdthesis}, where three samples (A, B, and X) are presented, A and B differ (e.g., distinct phonemes), and X matches either A or B. The ABX error rate is the proportion of incorrect classifications. We consider two setups: (1) Context-Independent (CI), where A and B are different phonemes in isolation, and (2) Context-Dependent (CD), where A and B are triphones to account for coarticulation. The ABX triples are extracted from the LibriSpeech test sets (clean \& other), available in abxLS, an evaluation task in the zero-resource-speech benchmark \cite{dunbar2021zero}. All models outperform the reference MFCC-based error rate, with DAC and HAC outperforming the baseline ST-10-HuB-en. Among the DAC variants, DAC-14-HuB-en performs better than DAC-10-HuB-en, likely due to its higher bitrate. Introducing a transformer encoder before the student VQ layer as in DAC-14-HuB-T-en further reduces ABX error. HAC models show slightly higher ABX errors than DAC possibly because juggling both phoneme-level ($\mathcal{L}_{\text{KD-Phn}}$) and lexical-level ($\mathcal{L}_{\text{KD-Lex}}$) losses introduces additional training complexity. We also report ABX error rates using HuBERT (HuB) and mHuBERT (mHuB) embeddings for reference. 

Figure \ref{fig:pnmi_multi} shows results for Phoneme Normalized Mutual Information (PNMI) \cite{hsu2021hubert}, which measures the mutual information between tokens and true phoneme labels, normalized by phoneme label entropy. Values range from 0 (no correspondence) to 1 (perfect alignment). We evaluate PNMI across eight MLS languages. The multilingual DAC-14-mHuB-T-L16 model achieves the highest average PNMI, followed by the HAC multilingual variants; ST-10-HuB-en ranks lowest. Consistent with the ABX results, adding a transformer encoder before the phonetic VQ layer (e.g., DAC-14-HuB-T-en vs.\ DAC-14-HuB-en) significantly boosts PNMI. English performance is strongest overall, reflecting its greater share of training data compared to other languages.

Figure \ref{fig:word_detects} explores how well the tokens align with word labels. We extract tokens from the ${\tt VQlex}$ layer for HAC models. For DAC and ST models, we use the phonetic VQ layer. Following \cite{Harwath2020Learning}, we compute the F1 score for each (word, token) pair to see if the discovered tokens capture lexical items. Figure~\ref{fig:word_detects} plots how many tokens achieve an F1 score above various thresholds. HAC models exhibit a clear advantage, producing substantially more word detectors than their DAC or ST counterparts. HAC-14-LaBSE-HuB-T-en attains the highest number of strong word detectors, leveraging LaBSE-based text embeddings for lexical-level KD. HAC-14-SAMU-HuB-T-en ranks a close second; notably, it does not require text transcripts to compute the KD loss because it uses SAMU, a language-agnostic semantic speech encoder. Multilingual HAC models perform slightly worse, as expected given that evaluation is on English words only.
\begin{figure}
    \centering
    \includegraphics[width=0.8\linewidth]{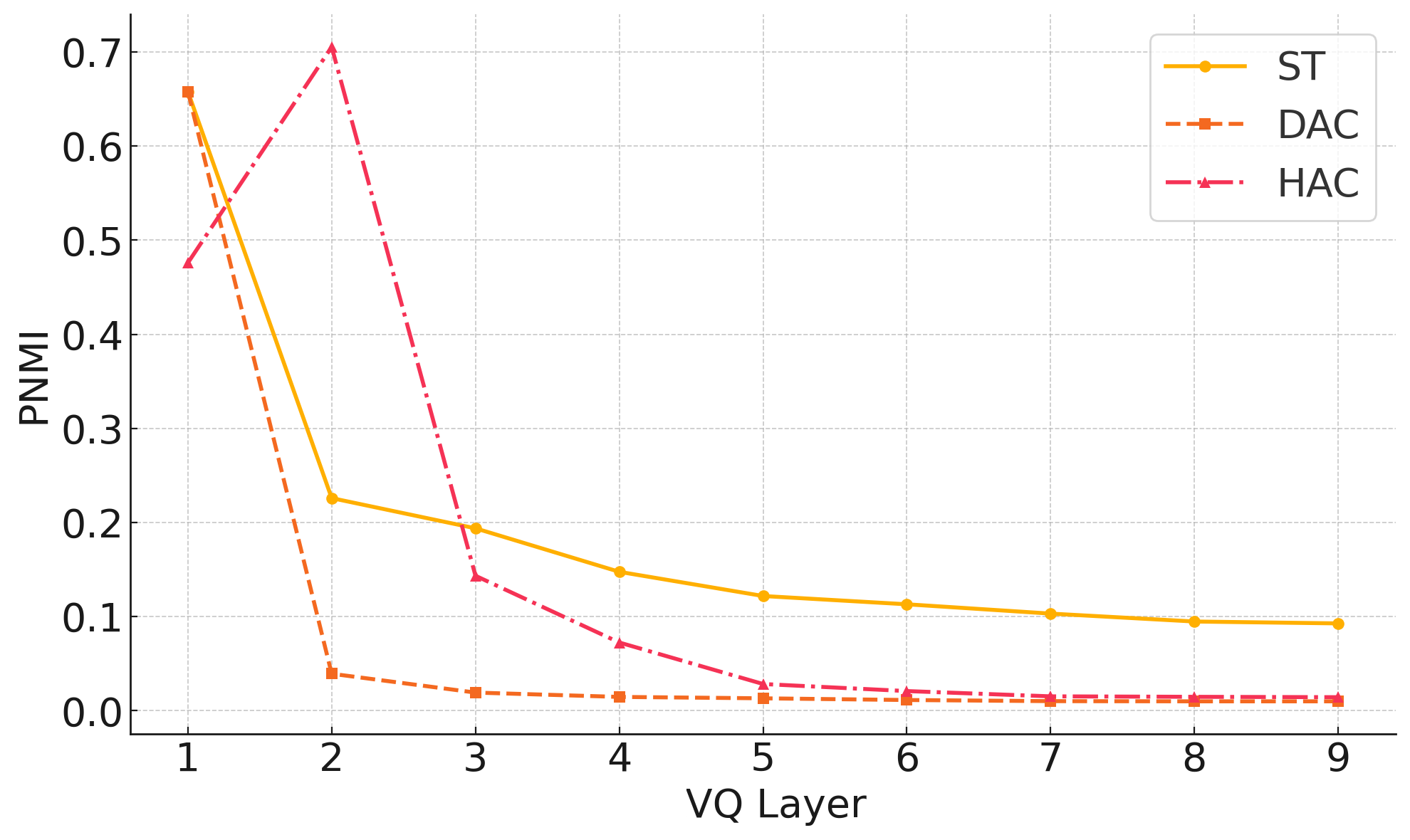}
    \vspace{-.1cm}
    \caption{Layer-wise Phoneme Normalized Mutual Information for different models.}
    \vspace{-.4cm}
    \label{fig:pnmi_layerwise}
\end{figure}
\begin{figure}
    \centering
    \includegraphics[width=0.8\linewidth]{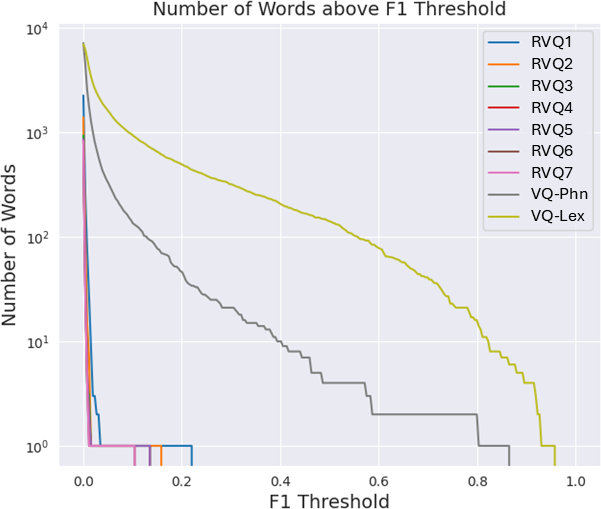}
    \vspace{-.1cm}
    \caption{F1 Scores of VQ tokens from different VQ layers of HAC when treated as word detectors.}
    \vspace{-.2cm}
    \label{fig:disent}
\end{figure}

Figures \ref{fig:pnmi_layerwise} and \ref{fig:disent} provide evidence of how different VQ layers in HAC encode distinct linguistic abstractions. 

Figure \ref{fig:pnmi_layerwise} compares layer-wise PNMI for HAC (HAC-10-SAMU-HuB-T-en), DAC (DAC-10-HuB-T-en), and ST (ST-10-HuB-T-en). VQ layers 1 and 2 in HAC refer to the lexical (attached to $\mathcal{L}_{\text{KD-Lex}}$) and phonetic (attached to $\mathcal{L}_{\text{KD-Phn}}$) VQ layers, respectively. In DAC and ST, VQ layer 1 refers to the phonetic layer. All models show partial disentanglement: phonetic layers yield higher PNMI, whereas other layers do not align with phoneme labels. Notably, HAC’s lexical VQ layer (Layer 1) also exhibits some phoneme alignment but remains less phonemically oriented than the phonetic layer. Overall, DAC and HAC achieve stronger disentanglement than ST.

Figure~\ref{fig:disent} shows correspondence between tokens from different layers of HAC and the word labels. We observe that the layer ${\tt VQlex}$ has significantly more codebook entries that act as word detectors compared to other layers. Some codebook entries of the $\tt VQphn$ layer act as word detectors, and virtually no codebook entries from the $\tt RVQacoust$ module of HAC act as word detectors.
\begin{table}[t]
    \centering
    \sisetup{
    detect-weight, 
    mode=text, 
    tight-spacing=true,
    round-mode=places,
    round-precision=2,
    table-format=1.2,
    table-number-alignment=center
    }
    \caption{Reconstruction metrics across different models.}
    \label{tab:recons}
    \vspace{-8pt}
    \setlength{\tabcolsep}{3pt}
    \resizebox{\columnwidth}{!}{
    \begin{tabular}{lSSSS}\toprule
        Model &  {Mel-D $\downarrow$} & {STFT-D $\downarrow$} & {SI-SDR [dB] $\uparrow$} & {ViSQOL $\uparrow$}\\\midrule
        ST & 0.64 & 1.42&6.24&3.89\\
        HAC & 0.58 & 1.37&7.42&4.34\\
        DAC & 0.55&1.34&7.82&4.50\\
        DAC (Orig.)&0.51&1.29&8.01&4.55\\
        \bottomrule
    \end{tabular}
    }
    \vspace{-.3cm}
\end{table}

Finally, Table \ref{tab:recons} compares the reconstruction quality of different models on LibriSpeech clean test set using standard reconstruction metrics (see Section 4.4 of \cite{kumar2023highfidelity} for details). Both DAC and HAC outperform the baseline ST and achieve reconstruction quality on par with DAC (Orig.) \cite{kumar2023highfidelity}, an RVQ-GAN trained solely on generative losses. For brevity, DAC refers to DAC-10-HuB-T-en, ST to ST-10-HuB-en, and HAC to HAC-10-SAMU-HuB-T-en throughout the table.

\section{Conclusions}
This paper introduced Hierarchical Audio Codec (HAC), a factorized RVQ-GAN framework that unifies acoustic, phonetic, and lexical token sets within a single model. Through dedicated knowledge distillation losses from speech-focused (HuBERT) and text-based (LaBSE) encoders, HAC learns complementary token groups at different levels of linguistic abstraction. Specifically, the acoustic tokens capture the fine-grained spectral details needed for natural-sounding speech reconstruction, while the phonetic tokens align closely with underlying phoneme sequences, and the lexical tokens detect word-level distinctions. Experiments demonstrate that this disentangled multi-level representation not only preserves high-fidelity audio quality but also offers strong linguistic and semantic interpretability across various languages. Overall, our results highlight the potential of factorized tokenization to bridge the gap between high-fidelity audio compression and linguistically rich speech representations.

{\noindent \bf Acknowledgements---}Dominik Klement was supported by Czech Ministry of Interior project No.\ VK01020132 "112". The work reported here was started at JSALT 2024, and supported by Johns Hopkins University. This project has received funding from the European Union’s Horizon 2020 research and innovation programme under the Marie Skłodowska-Curie grant agreement No. 101007666. This work was granted access to the HPC resources of IDRIS under the allocations A0161014876 and A0181012527 made by GENCI.
\bibliographystyle{IEEEtran}
\bibliography{template}

\end{document}